\documentclass{tlp}

\usepackage{booktabs}
\usepackage{dsfont}
\usepackage{wrapfig} 
\usepackage{amsmath}
\usepackage{graphicx}
\usepackage{multirow}
\usepackage{multicol}
\usepackage{xspace}
\usepackage{xcolor}
\definecolor{linkcolor}{RGB}{52,59,144}
\usepackage[hidelinks, colorlinks=true, allcolors=linkcolor]{hyperref}
\usepackage{url}

\usepackage{comment}    
\usepackage{hyperref}

\usepackage{listings}

\usepackage[most]{tcolorbox}
\usepackage{enumitem}

\definecolor{noteColor1} {rgb}{  1,  0,  0} %
\definecolor{noteColor2} {rgb}{  1,0.5,  0} %
\definecolor{noteColor13}{rgb}{0.9,0.2,  0} 
\definecolor{noteColor3} {rgb}{0.8,0.8,  0}
\definecolor{noteColor4} {rgb}{  1,  0,0.6}
\definecolor{noteColor5} {rgb}{  1,  0,  1} %
\definecolor{noteColor6} {rgb}{  0,0.8,  0} %
\definecolor{noteColor7} {rgb}{  0,0.8,0.6} %
\definecolor{noteColor8} {rgb}{  0,0.8,0.8} %
\definecolor{noteColor9} {rgb}{0.5,0.8,  0} %
\definecolor{noteColor10}{rgb}{  0,  0,  1} %
\definecolor{noteColor11}{rgb}{0.5,  0,  1} %
\definecolor{noteColor12}{rgb}{  0,0.5,  1} %

\definecolor{maroon}{rgb}{0.5,0,0}
\definecolor{darkgreen}{rgb}{0,0.5,0}
\definecolor{mildgreen}{rgb}{0,0.8,0}
\definecolor{lightblue}{rgb}{0.3,0.3,1}
\definecolor{lightgrey}{rgb}{0.97,0.97,0.97}
\definecolor{grey}{rgb}{0.65,0.65,0.65}

\lstnewenvironment{lstlistingwolinenum}{
    \lstset{
        basicstyle=\ttfamily\footnotesize,
        frame=lrtb,
        breaklines=true
    }
}{}
\lstnewenvironment{lstlistingwlinenum}{
    \lstset{
        basicstyle=\ttfamily\footnotesize,
        frame=lrtb,
        breaklines=true,
        numbers=left,
        numbersep=0.5em,
        xleftmargin=1.5em,
        framexleftmargin=1.5em,
        stepnumber=1
    }
}{}

\usepackage{adjustbox}
\usepackage{relsize}
\definecolor{PrologPredicate}{RGB}{0,0,200}
\definecolor{PrologVar}      {RGB}{145,032,039}
\definecolor{PrologComment}  {RGB}{0,170,0}
\definecolor{PrologOther}    {rgb}{0.2,0.2,0.2}
\definecolor{PrologString}   {RGB}{070,120,200}
\usepackage{listings} 
\usepackage{textcomp}
\newcommand{\code}{\lstinline[style=MyInline]}
\lstdefinestyle{MyInline}
{
  basicstyle = \relsize{-0.5}\ttfamily\color{PrologPredicate},
  escapechar = @,
  escapeinside = {-<}{>-},
  breaklines = true,
  breakatwhitespace=true,
  upquote = true,
  literate =
  {?-}{{?-\,}}3
  {:-}{{:-\,}}3
  {.=.}{{\,\#=\,}}3
  {.<.}{{\,\#<\,}}3
  {.>.}{{\,\#>\,}}3
  {.=<.}{{\,\#=<\,}}4
  {.>=.}{{\,\#>=\,}}4
}
\lstdefinestyle{MySCASP}
{
    numbersep=0.5em,    
    xleftmargin=0.35cm,  
    numberstyle=\tiny,
    numbers=left,
    stepnumber=1,
  mathescape = true,
  escapechar = @,
  escapeinside = {-<}{>-},
  keywords = {},
  upquote = true,
  basicstyle = \ttfamily\relsize{-.5}\color{PrologPredicate},
  basewidth = 0.47em,
  moredelim = {*[s][\color{PrologVar}]{(}{)}},
  moredelim = {*[s][\color{PrologString}]{'}{'}},
  moredelim = {*[s][\color{PrologOther}]{:-}{.}},
  commentstyle = \mdseries\color{PrologComment},
  escapebegin=\color{PrologVar},
  morecomment=[l]\%,
  literate     =
  {|}{{|}}2
  {.=.}{{\,\#=\,}}3
  {.<.}{{\,\#<\,}}3
  {.>.}{{\,\#>\,}}3
  {.=<.}{{\,\#=<\,}}4
  {.>=.}{{\,\#>=\,}}4
}

\usepackage{ntheorem}

{
\theoremindent=.5cm
\theoremheaderfont{\kern-.5cm\normalfont\slshape}
\theorembodyfont{\normalfont}
\theoremseparator{.}
\theoremprework{\setlength\parindent{0cm}}

}

\newcommand{\scasp}[0]{s(CASP)}
\newcommand{\clingo}[0]{\textsc{Clingo}}
\newcommand{\hilite}[0]{\textsc{HiLiTE}}

\newcommand{\myurl}[1]{\href{http://#1}{\nolinkurl{#1}}}
\newcommand{\version}{0.24.04.04\xspace}

\newtcolorbox{reqCodeBox}[3][0.6em]{%
    title={#3},
    colframe=black!75,
    enhanced,
    use color stack,
    breakable,
    left=2pt, right=0pt, top=-6pt, bottom=-6pt, boxsep=0pt, boxrule=0.25pt, arc=2pt,
    bottomsep at break=6pt,
    topsep at break=6pt,
    coltitle=white, 
    detach title,
    overlay unbroken and first={
    \node[xshift=-#2/2,yshift=-#1] at (frame.north east) {%
        \begin{tcolorbox}[enhanced, coltitle=white, colback=black!75, width=#2, left=0pt, right=0pt, top=1pt, bottom=1pt, boxsep=1pt, boxrule=0pt, arc=2pt]\textcolor{white}{#3}\end{tcolorbox}
        };
    }
}
\newtcolorbox{codeBox}{%
    colframe=black!75,
    enhanced,
    use color stack,
    breakable,
    left=2pt, right=0pt, top=-6pt, bottom=-6pt, boxsep=0pt, boxrule=0.25pt, arc=2pt,
    bottomsep at break=6pt,
    topsep at break=6pt,
    coltitle=white, 
    detach title,
}
\newtcolorbox{enumBox}{%
    colframe=black!75,
    enhanced,
    use color stack,
    breakable,
    left=2pt, right=2pt, top=0pt, bottom=0pt, boxsep=2pt, boxrule=0.25pt, arc=2pt,
    bottomsep at break=0pt,
    topsep at break=0pt,
    coltitle=white, 
    detach title,
}

\begin{document}

\jnlPage{X}{Y}
\jnlDoiYr{2024}
\doival{10.1017/xxxxx}


\title[
  Early Validation of High-level System Requirements with EC and ASP        
]
{ 
  Early Validation of High-level System Requirements with Event Calculus and Answer Set Programming%
  \thanks{We are grateful to anonymous reviewers for their insightful comments and suggestions for improvement. The Czech team was supported by the project 23-06506S of the Czech
  Science Foundation and the FIT BUT internal project FIT-S-23-8151. Joaquin Arias was supported by grant VAE (TED2021-131295B-C33) funded by MCIN/AEI/10.13039/501100011033 and by the ``European Union NextGenerationEU/PRTR'', by grant COSASS (PID2021-123673OB-C32) funded by MCIN/AEI/ 10.13039/501100011033 and by ``ERDF A way of making Europe''. Gopal Gupta
  was partially supported by US NSF Grants IIS 1910131 and grants from
  industry through the UT Dallas Center for Applied AI and Machine Learning.}    
}
\lefttitle{
  O. Va\v{s}\'{i}\v{c}ek et al.         
}

\begin{authgrp}

    \author{%
               \sn{Va\v{s}\'{i}\v{c}ek}, \gn{Ond\v{r}ej}$^1$    \hfill      
               \sn{Arias},               \gn{Joaquin}$^2$       \hfill      
               \sn{Fiedor},              \gn{Jan}$^{1,7}$       \hfill     
               \sn{Gupta},               \gn{Gopal}$^3$                  \\
        \hfill \sn{Hall},                \gn{Brendan}$^4$       \hfill     
               \sn{K\v{r}ena},           \gn{Bohuslav}$^1$      \hfill     
               \sn{Larson},              \gn{Brian}$^5$         \hfill ~ \\
        \hfill \sn{Varanasi},            \gn{Sarat Chandra}$^6$ \hfill     
               \sn{Vojnar},              \gn{Tom\'{a}\v{s}}$^{1,8}$ \hfill ~   
    }
    \vspace{.5em}
    \affiliation{%
        $^1$Brno University of Technology, CZ \hspace{.2cm}  %
        $^2$Universidad Rey Juan Carlos, ES \hspace{.2cm}    %
        $^3$UT Dallas, USA \hspace{.2cm}         %
        $^4$Ardent Innovation Labs, USA \hspace{.2cm}
        $^5$Multitude Corp., USA\hspace{.2cm}                        
        $^6$GE Aerospace Research, USA \hspace{.2cm}       %
        $^7$Honeywell International s.r.o., CZ \hspace{.2cm}        %
        $^8$Masaryk University, CZ
    }
\end{authgrp}

\history{\sub{13.05.2024;} \rev{16.07.2024;} \acc{06.08.2024}}

\maketitle

\begin{abstract}
This paper proposes a~new methodology for early validation of high-level requirements on cyber-physical systems with the aim of improving their quality and, thus, lowering chances of specification errors propagating into later stages of development where it is much more expensive to fix them.
The paper presents a~transformation of a~real-world requirements specification of a
medical device---a~PCA pump---into an Event Calculus model that is then evaluated using Answer Set Programming and the \scasp{} system.
The evaluation under \scasp{} allowed deductive as well as abductive reasoning about the specified functionality of the PCA pump on the conceptual level with minimal implementation or
design dependent influences,
and led to fully-automatically detected nuanced violations of
critical safety properties. 
Further, the paper discusses scalability and non-termination challenges that had
to be faced in the evaluation and techniques proposed to (partially) solve them.
Finally, ideas for improving \scasp{} to overcome its evaluation limitations that still persist as well as to
increase its expressiveness are presented.
\end{abstract}

\begin{keywords}
Requirements Validation, Event Calculus, Answer Set Programming, \scasp{}
\end{keywords}


\section{Introduction and Background}
\label{sec:intro}


Early validation of specifications describing requirements placed on cyber-physical systems (CPSs) under development is essential to avoid costly errors in later stages of the development, especially when the systems undergo certification.
However, there is a lack of suitable automated tools and techniques for this purpose.
A~crucial need here is that of a small semantic gap between the requirements and the formalism used to model them for the purposes of validation.
A~larger semantic gap makes it more difficult to transform the requirements into a model, and, most importantly, any validation on such a model drifts away from validating the requirements themselves and closer to validating that particular model---influenced by design and implementation decisions.
Furthermore, when reasoning about
safety-critical systems, it is necessary---both from engineering and legal points of view---that the tools used must be able to explain the result of the validation.

As described by~\cite{mueller_book-fixed}, Event Calculus (EC) is a formalism suitable for commonsense reasoning.
The semantic gap between a requirements specification and its EC encoding is near-zero because its semantics follows how a~human would think of the requirements.
%
%
Using Answer Set Programming (ASP) and the \scasp{} system for goal-directed reasoning in EC,
the work \cite{gupta-train} has demonstrated the versatility of EC for modeling and reasoning about CPSs while providing explainable results.
However, the CPS presented is still a rather toy system only.
\\

%
In this work, we develop a model, presented in Section~\ref{sec:model-patient-bolus}\footnote{For the reader's convenience, the files described/used in the paper are available and linked to a~GitHub repository available at \url{https://github.com/ovasicek/pca-pump-ec-artifacts/}.}, of the core operation of the PCA pump by~\cite{pcapump-paper}---a~real safety-critical device.
The model operates in a~way similar to an early prototype of the system and, thus, can be used to reason about its behaviour.
However, due to the nature of EC, the behaviour of the model is very close to the behaviour described by the requirements themselves.
This allows us to reason about the requirements without tainting the reasoning by implementation or design decisions, which would be necessary when using lower-level models or a~physical~prototype.

Our work has resulted in the discovery of a~number of issues in the PCA pump specification.
%
Using automated reasoning, we were able to discover inconsistencies between the requirements specification and the use cases and exception cases based on which the requirements were created (Section~\ref{sec:uc2}).
%
We were further able to detect a~safety property violation which can lead to an overdose of the patient (Section~\ref{sec:ec13}).
Such discoveries could otherwise occur much later in the development process.
We have discussed and confirmed the issues with the authors of the specification.

We present a~number of challenges encountered during the translation of the requirements to EC encoded in s(CASP) and during the subsequent evaluation, based on deductive as well as abductive reasoning, which was often too costly or non-terminating.
We have applied and, in multiple cases, also newly developed various techniques that helped us resolve many of these challenges.
These include extensions of the axiomatization of the EC and special ways of translating certain parts of the specifications (Sections~\ref{sec:imp-ec-axioms} and~\ref{sec:self-ending-trajectories}), which we believe may be inspiratory when modelling and evaluating other systems too.
Further, we present an original approach to abductive reasoning with incrementally refined abduced values in order to assure consistency of the abduced values whenever abduction on the same value is used multiple times in the reasoning tree (Section~\ref{sec:incremental-abduction}). 
Next, we proposed a~mechanism for caching predicate evaluations (failure-tabling and tabling of ground sub-goal success) that was added into \scasp{} as a prototype leading to a~significant increase in performance (Section~\ref{sec:cache}).
We also describe a way of separating the reasoning about the trigger and the effect of certain complexity-inducing triggered events into multiple reasoning runs where each run produces new facts to be used in the subsequent ones (Section~\ref{sec:multi_run_approach}), which reduces their performance impact.
Finally, we propose two new lines of work (Section~\ref{sec:conclusions}), including a more systematic treatment of caching.

\paragraph*{Related Work}
%
Above, we have emphasized the suitability of EC for reasoning about requirements specifications due to its low semantic gap against them.
In comparison, the semantics of automata-based approaches, such as timed automata in UPPAAL by~\cite{timed-automata-uppaal}, require one to ``design" explicit states and transitions, and may lead to decomposition of the system into sub-systems each with their own automaton.
Current industrial model-based engineering approaches, such as those based, e.g., on Matlab Simulink models and tools like \hilite{} by~\cite{hilite}, are only suitable for validation of low-level requirements.
This is due to the low-level nature of the models they use, especially when automated generation of code from the models is required.

Apart from the EC-based approach
introduced by \cite{gupta-train}, which this work builds upon, 
there are other ones which aim to target automated validation of high-level requirements put on CPSs.
The work by \cite{ge-assert} is based on ontologies and uses theorem proving, which traditionally requires significant manual work.
The work by \cite{cea-glossaries-and-process-algebras} is based on transforming CPS specifications from templated-English
into process algebras extended with real-time aspects, however, no continuous variables (apart from time) are dealt with and no experimental results are presented, which makes it difficult to judge the scalability of this approach. A~more detailed comparison of the approaches is an interesting future work.

A transformation of a CPS into EC was considered in the already mentioned work by \cite{gupta-train}.
However, it considered a simple Train-Gate-Controller system only.
We expand on that work by transforming a~more complex, real-life specification
of the PCA pump, which has led to the discovery of a number of issues that did not manifest in the simpler system.
We tackle the issues by introducing techniques for avoiding non-termination and improving performance when reasoning about EC in \scasp{}.
In addition, we further propose a~way to check consistency between levels of the specification and to leverage abductive reasoning.

Finally, we note that there are of course other ASP solvers than the \scasp{} system we used.
Notably, grounding-based ASP solvers, such as \clingo{} by~\cite{clingo-paper}, are well known.
However, such solvers are, unfortunately, not suitable for reasoning about fluents with large or continuous value domains due to the explosion in the grounding and a need to discretize the time.
In our preliminary attempts at modelling the PCA pump using Clingo, the solver could only reason about narratives with very restricted value domains of all fluents and with a small number of large time steps without running out of memory on a machine with 64\,GB of RAM while taking close to an hour of execution time.
Further, the need to discretize time requires approximation of time steps for reaching exact 
values of continuous fluents during periods of continuous change, which can lead to inaccurate behaviour of the model.
In comparison, the grouding-free nature of \scasp{} supported by constraint solvers allowed us to reason over continuous time, and increasing the value domain of a~fluent typically did not affect the solution time needed.
Consequently, \scasp{} was able to reason about the same narratives as our preliminary Clingo model using only up to 50\,MB of memory and around 5\,minutes of execution time.
A~thorough comparison of the solvers is out of scope of this work.
Some comparisons have already been made by~\cite{scasp-iclp2018} and by \cite{gupta-train}. 
A~very interesting future work would be revisiting the PCA pump model using
Clingo once sufficient advancements are made in avoiding the explosion in the grounding size, especially since Clingo does not suffer from non-termination issues, which make things much more complicated in s(CASP).

\section{Preliminaries}
\label{sec:background}

This section describes (i) s(CASP), a~goal-directed implementation of Answer Set Programming (ASP) with Constraints, and the Event Calculus (EC), a~formalism for reasoning about events and change, and
(ii) an open source Patient-Controlled Analgesia (PCA) pump specification, which 
we use as a~real use case.

\subsection{The \scasp{} System  and the Event Calculus}
\label{sec:scasp-asp-ec}

The s(CASP) system, presented by~\cite{scasp-iclp2018}, extends the
expressiveness of ASP systems, based on the stable
model semantics by~\cite{gelfond88:stable_models-fixed}, by including
predicates, constraints among non-ground variables, uninterpreted
functions, and, most importantly, a~top-down, query-driven execution
strategy.
These features make it possible to return answers with non-ground
variables (possibly including constraints among them) and to compute
partial models by returning only the fragment of a~stable model that
is necessary to support the answer to a~given query.
Answers to all queries can also include the full proof tree, making them fully explainable.

In s(CASP), thanks to the constructive negation, %
\code|not p(X)| can return bindings for~\code{X} for which the call
\code{p(X)} would have failed.  Thanks to the interface of s(CASP)
with constraint solvers, sound non-monotonic reasoning with
constraints is possible.

Like other ASP implementations and unlike Prolog, s(CASP) handles
non-stratified negation and returns the corresponding (partial) stable
models, e.g., for the program \mbox{\code{p :- not q.  q :- not p.}}, under
the stable model semantics there are two possible models for this \textit{even loop} (\cite{asp}), with either
\code{p} or \code{q} being true.
Even loops are used in \scasp{} to implement abductive reasoning via the \code{#abducible} directive, where we automatically search for suitable values of the predicates in the corresponding even loop so we can satisfy the main query.
We use abduction in Section~\ref{sec:ec13} to detect a~violation of a~critical safety property in the PCA pump requirements.
\\


The Event Calculus (EC) is a~formalism for reasoning about events and
change by~\cite{mueller_book-fixed}, of which there are several
axiomatizations.  There are three basic concepts in EC:
\emph{events}, \emph{fluents}, and \emph{time points}:
(i) an event is an action or incident that may occur in the world, e.g., the dropping of a~glass by a~person is an event,
(ii) a~fluent is a~time-varying property of
the world, such as the altitude of a~glass,
(iii) a~time point is an instant of time.
Events may happen at a~time point; fluents have a~truth value at any
time point
, and these truth values are subject to
change upon an occurrence of an event.  In addition, fluents may have
quantities associated with them as parameters, 
which change discretely via events or continuously over time via trajectories.

E.g., the event of \emph{dropping} a~glass initiates the fluent that
captures that the glass is \emph{falling}, which enables a~trajectory that determines the decreasing value of a~fluent that represents the glass's \emph{height} above the ground, and the event of \emph{catching} a~glass terminates the fluent that the glass is \emph{falling}, which disables the trajectory.
An EC description consists of a~universal theory and a~domain narrative
(see the book by~\cite{mueller_book-fixed} for details).
The theory is a~conjunction of EC axioms, e.g.,
axiom BEC6 states that a~fluent $f$ is true at a~time $t_2$ if it is
initiated by some event $e$ occurring at some earlier time $t_1$ and
it is not stopped between $t_1$ and $t_2$:
$$  \mathit{HoldsAt}(f, t_2) ~~\leftarrow~~ \mathit{Happens}(e, t_1) \land \mathit{Initiates}(e, f, t_1) \land 
     t_1 < t_2 \land \lnot \mathit{StoppedIn}(t_1, f, t_2). $$

\noindent
The domain narrative consists of the causal laws of the domain and
the known events and fluent properties.  \cite{mueller_book-fixed}, in his book in Example 14, reasons about turning a~light switch on and off using the event
$\mathit{Happens}(e,t) ~\equiv~ (e=\mathit{TurnOn} ~\land~ t=1/2)$ 
  \mbox{$\lor~ (e=\mathit{TurnOff} ~\land~ t=4)$}
that states that the $\mathit{TurnOn}$ event will happen exclusively at time $t=1/2$
and that $\mathit{TurnOff}$ will happen exclusively at $t=4$. 
\\

Two key factors contribute to the s(CASP)'s ability to model EC: the preservation of non-ground variables during the
execution and the integration with constraint solvers.
Using the translation rules introduced by~\cite{arias-ec2022}, one can translate the BEC axioms by~\cite{mueller_book-fixed} into s(CASP) programs
that follow the logic programming convention: constants and predicate
symbols start with a~lowercase letter, variables start with an
uppercase letter, and constraints can be written
with a~prefix (\texttt{\#<}). For example, the
BEC6 axiom and events are translated as: 

\begin{codeBox}\vspace{2pt}
\setlength{\columnsep}{30mm}
\begin{multicols}{2}
\begin{lstlisting}[style=MySCASP]
holdsAt(F, T2) :- T1.<.T2, initiates(E, F, T1),
  happens(E, T1), not stoppedIn(T1, F, T2).
happens(turn_on, 1/2).
happens(turn_off, 4).
\end{lstlisting}
\end{multicols}
\end{codeBox}

    \subsection{The Patient-Controlled Analgesia (PCA) Pump Project}
    \label{sec:pati-contr-analg}
    The open PCA Pump Project,
    introduced by~\cite{pcapump-paper} and available at \url{https://openpcapump.santoslab.org/}, provides a~full set of realistic artifacts used in the development process of a~\emph{Patient-Controlled Analgesia Infusion Pump}, which is a~safety-critical medical device.
    The artifacts were created at the behest of the US Food and Drug Administration to provide an open-source example of model-based systems engineering for industry, and a~subject matter for researchers.
    The primary function of the device is to automatically and safely deliver the appropriate amount of pain-relief drugs to a~patient via infusion into their bloodstream.
    The pump needs to do so without delivering an amount that would harm the patient, it needs to notify clinicians about hazards, and it needs to maintain safe operation even when failures occur or when hazards are detected.
    The delivery parameters, such as drug flow rates or maximum safe doses, are either prescribed by a~physician or specified in a~drug library.
    
    In this paper, we use version 1.0.0 of the PCA specification
    (\href{https://github.com/ovasicek/pca-pump-ec-artifacts/blob/master/Open-PCA-Pump-Requirements.pdf}{Open-PCA-Pump-Requirements.pdf}
    in GitHub, which we reference as
    [\href{https://ovasicek.github.io/pca-pump-ec-artifacts/Open-PCA-Pump-Requirements.pdf}{PCA
    page~N}] in the following when referring to page N).
    The PCA pump delivers drug using four different types of delivery:
    (i)~\emph{Basal delivery} is the baseline which delivers drug using a~small flow rate during normal operation.
    It is the initial type of delivery after starting the pump.
    (ii)~\emph{Patient-requested bolus} is an extra dose which can be requested by the patient via a~button. 
    Upon a~valid request, the PCA pump delivers a~prescribed amount of drug called \emph{VTBI} (volume-to-be-infused) using a~higher flow rate in addition to the baseline basal flow rate, and then returns to basal delivery.
    (iii)~\emph{Clinician-requested bolus} is a~second, similar, extra dose of the same VTBI spread over many minutes.
    It differs in that it can only be requested by a~clinician and that they can select a~duration for the bolus. 
    (iv)~\emph{KVO delivery} (Keep-Vein-Open) is an emergency delivery with the smallest flow rate to prevent clotting of the needle in response to certain alarms.

    The specification defines a~number of alarms and how the PCA pump should respond to them.
    Most of the alarms are related to hardware failures or physical issues detected by sensors, while others are raised by the logic of the PCA pump, e.g., to prevent an overdose of the patient.

\section{Modelling the PCA Pump Requirements in EC under s(CASP)}
\label{sec:model-patient-bolus}

The requirements are specified in unconstrained natural language, which makes automated processing difficult, and so their transformation to an EC model was done manually.
An automated transformation from more structured requirements is part of our future work.
We have modeled the PCA pump based on Chapter \emph{II. Requirements} [\href{https://ovasicek.github.io/pca-pump-ec-artifacts/Open-PCA-Pump-Requirements.pdf#page=64}{PCA page~54}].
Our main focus was on the core functionality of the PCA pump, defined in Section \emph{5 PCA Pump Function} [\href{https://ovasicek.github.io/pca-pump-ec-artifacts/Open-PCA-Pump-Requirements.pdf#page=65}{PCA page~55}], and we omitted a portion of requirements stated in other sections (e.g., non-essential features and physical properties).
We do not cover the entire transformation here due to space limitations.
Below, we demonstrate it on several representative examples.
All source files are available at \href{https://github.com/ovasicek/pca-pump-ec-artifacts/}{https://github.com/ovasicek/pca-pump-ec-artifacts/},
general principles of modeling using EC are explained by~\cite{mueller_book-fixed}, and a~similar transformation of the Train-Gate-Controller has been shown by~\cite{gupta-train}.

The centerpiece of the PCA pump is the total amount of drug that has been delivered. 
We represent it by a~continuous fluent \code{total_drug_delivered(X)}. 
Its value is constant while the pump is stopped and it changes gradually at a~given rate while the pump is running.
The gradually changing value is given by the chosen type of drug delivery---pump stopped (no delivery), basal, patient bolus, clinician bolus, or KVO (Section~\ref{sec:pati-contr-analg}).
Each of the delivery types needs to be represented by an EC trajectory and the logic of the PCA pump then determines which trajectory is active at what time.
\\


    We demonstrate the transformation on requirements defined for the delivery of a~\emph{patient-requested bolus} [\href{https://ovasicek.github.io/pca-pump-ec-artifacts/Open-PCA-Pump-Requirements.pdf#page=65}{PCA page~55}] (implemented in \href{https://github.com/ovasicek/pca-pump-ec-artifacts/blob/master/model_sources/04-patient_bolus_trajectory.pl}{04-patient\_bolus\_trajectory.pl}).
    Other delivery modes were transformed in a~similar fashion\footnote{
        See \href{https://github.com/ovasicek/pca-pump-ec-artifacts/blob/master/model_sources/04-basal_delivery_trajectory.pl}{04-basal\_delivery\_trajectory.pl}, \href{https://github.com/ovasicek/pca-pump-ec-artifacts/blob/master/model_sources/04-clinician_bolus_trajectory.pl}{04-clinician\_bolus\_trajectory.pl}, \href{https://github.com/ovasicek/pca-pump-ec-artifacts/blob/master/model_sources/04-kvo_delivery_trajectory.pl}{04-kvo\_delivery\_trajectory.pl}, \href{https://github.com/ovasicek/pca-pump-ec-artifacts/blob/master/model_sources/04-pump_state.pl}{04-pump\_state.pl}, and other relevant files at \href{https://github.com/ovasicek/pca-pump-ec-artifacts/}{https://github.com/ovasicek/pca-pump-ec-artifacts/}.
    }.
    \begin{tcolorbox}[enhanced, breakable, left=0pt, right=2pt, top=3pt, bottom=3pt, boxsep=0pt, boxrule=1pt, arc=2pt]\small 
    \begin{itemize}
        \item [\textbf{R1:}] Upon patient's press of the PCA pump's patient-button, a~prescribed bolus volume-to-be-infused, VTBI, of the drug loaded in the pump is delivered to the patient.
    \end{itemize}
    \end{tcolorbox}
    %
    \noindent R1 introduces the patient bolus delivery mode in general.
    We define a~fluent to represent the delivery state, and its start/end events and their effects.
    Then, we define a~trajectory to determine the value of the \code{total_drug_delivered(X)} fluent while this delivery is active.%
    %
    \begin{reqCodeBox}[0.55em]{2em}{R1a}
    \begin{lstlisting}[style=MySCASP]
fluent(patient_bolus_delivery_enabled).
event(patient_bolus_delivery_started).   event(patient_bolus_delivery_stopped).
initiates(patient_bolus_delivery_started, patient_bolus_delivery_enabled, T).
terminates(patient_bolus_delivery_stopped, patient_bolus_delivery_enabled, T).

trajectory(patient_bolus_delivery_enabled, T1, total_drug_delivered(Total), T2) :-
  basal_and_patient_bolus_flow_rate(FlowRate),
  holdsAt(total_drug_delivered(StartTotal), T1),
  Total .=. StartTotal + ((T2 - T1) * FlowRate).
    \end{lstlisting}
    \end{reqCodeBox}
    \noindent And finally, we define that the bolus ends automatically once it delivers the full VTBI.
    This is represented by a~\code{patient_bolus_completed} event and its trigger rule.
    \begin{reqCodeBox}[0.55em]{2em}{R1b}
    \begin{lstlisting}[style=MySCASP, firstnumber=9]
event(patient_bolus_completed).
happens(patient_bolus_completed, T2) :-   initiallyP(vtbi(VTBI)),
  holdsAt(patient_bolus_drug_delivered(VTBI), T2).
happens(patient_bolus_delivery_stopped, T) :- happens(patient_bolus_completed, T).

fluent(patient_bolus_drug_delivered(X)).
trajectory(patient_bolus_delivery_enabled,T1, patient_bolus_drug_delivered(X),T2):-
  patient_bolus_only_flow_rate(FlowRate),
  X .=. (T2 - T1) * FlowRate.
    \end{lstlisting}
    \end{reqCodeBox}
    %
    \noindent The \code{patient_bolus_completed} event triggers once the amount of drug delivered via the bolus delivery rate reaches VTBI.
    This value is represented by a~new fluent and its trajectory, which allow easier tracking of the progress of the bolus by counting its value from zero.
    The new fluent is not affected by any events, it is given by the trajectory only.

    \begin{tcolorbox}[enhanced, breakable, left=0pt, right=2pt, top=3pt, bottom=3pt, boxsep=0pt, boxrule=1pt, arc=2pt]\small 
    \begin{itemize}
        \item [\textbf{R2:}] A~patient-requested bolus shall be delivered at its prescribed rate, $F_{bolus}$, in addition to the basal flow rate, $F_{basal}$, but no more than the max. flow rate for the pump, $F_{max}$.
    \end{itemize}
    \end{tcolorbox}
    %
    \noindent For R2, we define a~predicate which computes the flow rate based on the values of system parameters which are represented using constant fluents.
    %
    \begin{reqCodeBox}[0.55em]{1.5em}{R2}
    \begin{lstlisting}[style=MySCASP]
basal_and_bolus_flow_rate(Cropped) :-   initiallyP(pump_flow_rate_max(Max)),
  initiallyP(patient_bolus_flow_rate(Bolus)),   initiallyP(basal_flow_rate(Basal)),
  Combined .=. Bolus + Basal,   min(Combined, Max, Cropped).
  
patient_bolus_only_flow_rate(BolusOnly) :-   basal_and_bolus_flow_rate(Cropped),
  initiallyP(basal_flow_rate(Basal)),   BolusOnly .=. Cropped - Basal.
    \end{lstlisting}
    \end{reqCodeBox}
    
    \begin{tcolorbox}[enhanced, breakable, left=0pt, right=2pt, top=3pt, bottom=3pt, boxsep=0pt, boxrule=1pt, arc=2pt]\small 
    \begin{itemize}
        \item [\textbf{R6:}] Any alarm stops patient-requested bolus delivery either halting pump or switching to KVO rate as defined in Table 4 [\href{https://ovasicek.github.io/pca-pump-ec-artifacts/Open-PCA-Pump-Requirements.pdf#page=69}{PCA page~59}].
    \end{itemize}
    \end{tcolorbox}
    %
    \noindent To implement R6, we trigger the occurrence of the end of the bolus when \code{any_alarm} happens (which is itself triggered by specific alarms) while the bolus is active.
    %
    \begin{reqCodeBox}[0.55em]{1.5em}{R6}
    \begin{lstlisting}[style=MySCASP]
happens(patient_bolus_halted, T) :-   happens(any_alarm, T),
  holdsAt(patient_bolus_delivery_enabled, T).
happens(patient_bolus_delivery_stopped, T) :-   happens(patient_bolus_halted, T).
    \end{lstlisting}
    \end{reqCodeBox}

    \begin{tcolorbox}[enhanced, breakable, left=0pt, right=2pt, top=3pt, bottom=3pt, boxsep=0pt, boxrule=1pt, arc=2pt]\small 
    \begin{itemize}
        \item [\textbf{R3:}]Patient-requested bolus shall not be delivered more often than a~prescribed minimum time between patient-requested bolus, $\Delta_{prb}$.
        \item [\textbf{R5:}] Patient-requested bolus shall not be delivered if infusing prescribed VTBI will exceed hard limits retrieved from the drug library for the volume of drug infused over a~period of time. Pump rate shall be reduced to KVO and a~max dose warning be issued.
    \end{itemize}
    \end{tcolorbox}
    %
    %
    \noindent
    R3 and R5 define cases when a~requested bolus should be denied.
    This can be implemented by making the occurrence of \code{patient_bolus_delivery_started} conditional.
    %
    \begin{reqCodeBox}[0.7em]{2.5em}{R3/5}
    \begin{lstlisting}[style=MySCASP]
happens(patient_bolus_delivery_started, T) :-
  happens(patient_bolus_requested, T),
  not_happens(patient_bolus_denied_too_soon, T),
  not_happens(patient_bolus_denied_max_dose, T).
    \end{lstlisting}
    \end{reqCodeBox}
    %
    \noindent R3 is implemented by checking if any bolus delivery was enabled in the past while too close to the current request.
    %
    \begin{reqCodeBox}[0.55em]{1.5em}{R3}
    \begin{lstlisting}[style=MySCASP, firstnumber=4]
happens(patient_bolus_denied_too_soon, T) :-
  happens(patient_bolus_requested, T),
  initiallyP(min_t_between_patient_bolus(MinGap)),
  TLast .<. T,   TLast .>. T - MinGap,
  holdsAt(patient_bolus_delivery_enabled, TLast).
    \end{lstlisting}
    \end{reqCodeBox}
    %
    \noindent R5 is implemented by checking the total amount of drug delivered within the max dose time window (e.g., in the last hour).
    The delivery mode needs to be changed accordingly when a~warning is triggered.
    The code below is simplified for space reasons, details can be found in \href{https://github.com/ovasicek/pca-pump-ec-artifacts/blob/master/model_sources/04-patient_bolus_trajectory.pl\#L32}{04-patient\_bolus\_trajectory.pl}. 
    %
    \begin{reqCodeBox}[0.55em]{1.5em}{R5}
    \begin{lstlisting}[style=MySCASP, firstnumber=7]
happens(patient_bolus_denied_max_dose, T) :-
  happens(patient_bolus_requested, T),
  initiallyP(vtbi_hard_limit_over_time(VtbiLimit, TimePeriod)),
  holdsAt(total_drug_delivered(CurrentTotal), T),
  TstartPeriod .=. T - TimePeriod,
  holdsAt(total_drug_delivered(TotalAtStartPeriod), TstartPeriod),
  TotalInPeriod .=. CurrentTotal - TotalAtStartPeriod,
  TotalInPeriod .>. VtbiLimit.
  
happens(max_dose_warning,T) :-   happens(patient_bolus_denied_max_dose, T).
happens(basal_stopped,T) :-   happens(patient_bolus_denied_max_dose, T),
  holdsAt(basal_delivery, T).
happens(kvo_started,T) :-   happens(max_dose_warning, T).
    \end{lstlisting}
    \end{reqCodeBox}

\section{Reasoning About the PCA Pump Requirements Using s(CASP)}
\label{sec:evaluation}

\newcommand{\EXC}{E{\smaller{}x}C}
The specification defines a~number of use cases (UC) and exception cases (\EXC{}) in Section~\emph{4~System Operational Concepts} [\href{https://ovasicek.github.io/pca-pump-ec-artifacts/Open-PCA-Pump-Requirements.pdf#page=23}{PCA page~13}].
The requirements and the UC/\EXC{}s should be mutually consistent.
We simulate the behaviour of UC/\EXC{}s using the EC model, which was created based on the requirements, with the expectation that the model should behave exactly as defined in the UC/\EXC{}s.
An example is shown in Section~\ref{sec:uc2}.
If the behaviour of the model is inconsistent with the UC/\EXC{}s, then we have produced evidence of the requirements being inconsistent with the UC/\EXC{}s (up to correctness of the transformation).
The capabilities of s(CASP) are not limited to simulating the behaviour of the pump but allow us to reason about its general properties.
In Section~\ref{sec:ec13}, we reason about preventing an overdose of the patient---a~critical safety property.
    
    \subsection{Validating Consistency of Use/Exception Cases and the Requirements}
    \label{sec:uc2}

    We show an example of using \scasp{} reasoning to simulate \emph{UC2: Patient-Requested Bolus} in order to validate its consistency with the requirements specification.
    %
    \begin{tcolorbox}[enhanced, breakable, left=7pt, right=2pt, top=3pt, bottom=3pt, boxsep=0pt, boxrule=1pt, arc=2pt, title=UC2: Patient-Requested Bolus]\small 
    \begin{itemize}[itemsep=0pt,labelsep=0.1em,leftmargin=2.2em]
        \item [\textbf{Pre:}]
            \begin{enumerate}[itemsep=0pt,labelsep=0.2em,leftmargin=1.1em]
                \item Steps 1 to 14 of Normal Operation Use Case (UC1) completed.
                \item Basal rate being infused.
                \item Prescribed minimum time between boluses has elapsed.
            \end{enumerate}
        \item [\textbf{Post:}]
            \begin{enumerate}[itemsep=0pt,labelsep=0.2em,leftmargin=1.1em]
                \item Resume basal rate infusion.
            \end{enumerate}
        \item [\textbf{Step:}]
            \begin{enumerate}[itemsep=0pt,labelsep=0.2em,leftmargin=1.1em]
                \item Patient presses bolus request button.
                \item Time since last bolus compared with prescribed min. time between boluses (\emph{see} \EXC{}1).
                \item If not too soon, begin infusing VTBI (\emph{unless} \EXC{}13: Maximum Safe Dose).
                \item After prescribed VTBI has been infused, resume basal rate infusion.
            \end{enumerate}
    \end{itemize}
    \end{tcolorbox}
    \noindent
    A~use case consists of pre-conditions, a~sequence of steps, and post-conditions.
    We create a~narrative of event occurrences based on the pre-conditions and input events from the steps.
    Then, we form a~query based on the post-conditions and triggered events from the steps.
    The below excludes the initialization of system parameters, e.g., that the VTBI is $1\,ml$ and the bolus flow rate is $1\,ml.min^{-1}$. 
    The implementation can be found in \href{https://github.com/ovasicek/pca-pump-ec-artifacts/blob/master/narratives_and_queries/uc2.pl}{uc2.pl} and \href{https://github.com/ovasicek/pca-pump-ec-artifacts/blob/master/model_utils/general_utils.pl\#L81}{general\_utils.pl} (utility predicates \code{holdsIn/3} and \code{holdsAfter/2}).
    %
    \begin{reqCodeBox}[0.55em]{2.3em}{UC2}
    \begin{lstlisting}[style=MySCASP]
happens(start_button_pressed,    60).                                % Pre  1
happens(patient_bolus_requested, 120).                               % Step 1
?- holdsIn(basal_delivery_enabled,                         60, 120), % Pre  2
   initiallyP(min_t_between_patient_bolus(MinT)), T1 .=. 120 - MinT, % Pre  3 
   not_holdsIn(patient_bolus_delivery_enabled,             T1, 120), % Pre  3  
   not_happens(patient_bolus_denied_too_soon,                  120), % Step 2
   not_happens(patient_bolus_denied_max_dose,                  120), % Step 3
   happens(patient_bolus_delivery_started,                     120), % Step 3
   initiallyP(vtbi(VTBI)),  happens(patient_bolus_completed,    T2), % Step 4
   holdsAt(patient_bolus_drug_delivered(VTBI),                  T2), % Step 4
   happens(basal_delivery_started,                              T2), % Step 4 
   holdsAfter(basal_delivery_enabled,                           T2). % Post 1
    \end{lstlisting}
    \end{reqCodeBox}
    %
    \noindent The occurrence times of input events are randomly chosen. 
    The model should behave according to the UC for any event times.
    In the future, we plan on allowing narratives with events at variable timepoints \code{T} (currently, fixed narratives are required).
    The above query succeeds, meaning that the model and the requirements are consistent with UC2.

\vspace{-0.2em} 
\subsubsection{Results of Experiments with Consistency Validation}
\label{sec:simulation-comparison}

We have simulated all relevant UCs and \EXC{}s from the PCA pump specification
on a~2.67\,GHz Xeon CPU, using at most 40\,MB of memory.
Selected representative results are shown in Table~\ref{t:simulation-queries},
the rest can be found in Appendix~\ref{apx:full-table} (directly included in this arXiv version).

Some of the cases appear in multiple variants of the narrative.
For instance, in UC3, a~clinician-requested bolus can be delivered uninterrupted
(UC3a) or it may be suspended by a~patient-requested bolus and resumed
afterwards (UC3b).
To save space, we aggregate the measurements of variants of the same case that
led to the same result.
All implementations can be found in the
\href{https://github.com/ovasicek/pca-pump-ec-artifacts/blob/master/narratives_and_queries/}{narratives\_and\_queries
folder}.
\\
\newcommand{\uclink}[2]{\href{https://github.com/ovasicek/pca-pump-ec-artifacts/blob/master/narratives_and_queries/#1.pl}{#2}}
        
All UCs were simulated successfully, but quite a~few \EXC{}s failed.
This has led to the discovery of a~number of issues in the specification, such
as inconsistencies in alarm responses or defined constants. 
In particular, Step 1 of \EXC{}7c [\href{https://ovasicek.github.io/pca-pump-ec-artifacts/Open-PCA-Pump-Requirements.pdf#page=44}{PCA
page~34}] says that an alarm should be raised if the
drug flow rate exceeds the prescribed rate \emph{for longer than 10\,seconds},
while the requirement R6.4.0(4)
[\href{https://ovasicek.github.io/pca-pump-ec-artifacts/Open-PCA-Pump-Requirements.pdf#page=68}{PCA
page~58}] defines \emph{1\,minute} instead.
Very similar issues were found in \EXC{}7e and other \EXC{}s.
Further, the second post-condition of all variants~of~\EXC{}7 expects
\emph{infusion to be halted}, but \emph{Table~4}
[\href{https://ovasicek.github.io/pca-pump-ec-artifacts/Open-PCA-Pump-Requirements.pdf#page=69}{PCA
page~59}] \mbox{requires a \emph{switch to KVO delivery}.}

        Cases UC3b and \EXC{}13a-c are significantly slower than the others due
        to their narratives containing multiple bolus requests (2--3), while the
        others only contain one or none.
        In general, we have observed the biggest (exponential) increase in
        execution time when increasing the number of bolus requests, i.e., the
        number of system input events.
        %
        %
        \EXC{}21 is also slower despite featuring no bolus requests due to its
        use of full reasoning about the level of the drug reservoir, which is
        discussed in Section~\ref{sec:multi_run_approach}.

\begin{table}
\centering
\small

  \caption{Results of simulation of relevant use cases and exception cases}
  \vspace{0.2em}

\begin{tabular}{lllcr}
  \toprule
                              & \textbf{Use Case Name}    & \textbf{Variant}         & \textbf{Result} & \textbf{Time (s)} \\ 
  \midrule 
  \uclink{uc2} {UC2}          & Patient-Requested Bolus   & no variants              & OK              &  3.17             \\ 
  \uclink{uc3a} {UC3a}        & Clinician-Requested Bolus & not suspend              & OK              &  2.84             \\ 
  \uclink{uc3b} {UC3b}        & Clinician-Requested Bolus & suspended and resumed    & OK              & 37.13             \\ 
  \midrule
  \uclink{ec7a} {\EXC{}7a-f}  & Over-Flow Rate Alarm      & defined in \EXC{} step 1 & FAIL            & 1.53-4.62         \\ 
  \uclink{ec13a}{\EXC{}13a}   & Maximum Safe Dose         & during basal delivery    & FAIL            & 25.62             \\ 
  \uclink{ec13b}{\EXC{}13b-c} & Maximum Safe Dose         & during each bolus        & OK              & 31.61-53.45       \\ 
  \uclink{ec21} {\EXC{}21}    & Reservoir Empty           & no variants              & OK              & 40.99             \\ 
  \midrule
\end{tabular} \label{t:simulation-queries}
\end{table}

    \subsection{Validating the Requirements wrt. General Properties}
    \label{sec:ec13}
    %
    We use \emph{\EXC{}13: Maximum Safe Dose} 
    as an example of reasoning about general properties of the system and, later, to demonstrate abductive reasoning capabilities of \scasp{}.
    \EXC{}13 defines that the pump should prevent an overdose of the patient by reducing the drug flow rate.
    This is a~general property, and so \EXC{}13 had to be implemented in three narratives based on whether the overdose would occur during (a)~basal delivery, (b)~a~patient-requested bolus, or (c)~a~clinician-requested bolus.
    The implementations can be found in \href{https://github.com/ovasicek/pca-pump-ec-artifacts/blob/master/narratives_and_queries/ec13a.pl}{ec13a.pl}, \href{https://github.com/ovasicek/pca-pump-ec-artifacts/blob/master/narratives_and_queries/ec13b.pl}{ec13b.pl}, and \href{https://github.com/ovasicek/pca-pump-ec-artifacts/blob/master/narratives_and_queries/ec13c.pl}{ec13c.pl}.
    For example, \EXC{}13b contains 3~patient-bolus requests while the max dose prescription is defined to allow 2.5~boluses in 4~hours.
    The implemented narrative consists of \code{happens(start_button_pressed,60)} and three instances of \code{happens(patient_bolus_requested,T)} for \code{T} equal to \code{300}, \code{340}, and \code{380}.
    The third bolus would cause an overdose if delivered.
    This overdose is prevented by the requirement R5.2.0(5) (discussed as R5 in Section~\ref{sec:model-patient-bolus}), according to which the bolus will be denied.
    A~similar measure is defined for a~clinician-requested bolus by R5.3.0(7) [\href{https://ovasicek.github.io/pca-pump-ec-artifacts/Open-PCA-Pump-Requirements.pdf#page=66}{PCA
page~56}].
    
    The query \code{?- vtbi_hard_limit_exceeded_at_T_by_X(T, X)} (implemented in \href{https://github.com/ovasicek/pca-pump-ec-artifacts/blob/master/model_utils/analysis_utils.pl#L14}{analysis\_utils.pl}) checks the amount of drug delivered within the max dose time window with the end of the window at time \code{T}.
    It succeeds if the maximum dose was exceeded and will return by how much via \code{X}.
    This query returns no models on \EXC{}13b meaning that an overdose did not happen at any time \code{T}.
    However, if we modify the narrative by changing the bolus request times to \code{100}, \code{140}, and \code{180} (switching from \EXC{}13b to \EXC{}13a), then the query succeeds with bindings \code{T #> 295,T #< 681/2} and \code{X #> 0,X #< 1/2}.
    This overdose happens during basal delivery 
    due to a~missing requirement (discussed in Section~\ref{sec:overdose}).

        \subsubsection{Utilizing Abductive Reasoning}
        \label{sec:abductive_reasoning}
        In order to detect the basal overdose issue in the previous section, we had to be ``lucky" enough to define a~narrative in which the violation manifests, in the same way as with regular testing.
        To address this, we utilize the abductive reasoning capabilities of \scasp{}.
        
        Ideally, we would like abduction to check whether an overdose can occur in some narrative without any prior restrictions. 
        However, this is currently not possible in \scasp{} due to non-termination issues related to reasoning in continuous time.
        Such abduction is part of our future work.
        Instead, we fix a skeleton of a narrative (i.e., a sequence of input events to happen) and abduce values of various parameters of the narrative.
        In particular, we abduce the overdose parameters of the PCA pump via the predicate  \code{initiallyP(vtbi_hard_limit_over_time(VtbiLimit, TimePeriod))}, i.e., we abduce both the max dose volume and the size of the max dose window, which allows the reasoner to explore a~broad spectrum of overdose scenarios despite being restricted to a~fixed narrative of event occurrences.
        We also apply restrictions on the abducible values in order to keep them meaningful, such as that the time period must be longer than the duration of a~single bolus and that the max dose volume must be big enough to fit a~full period of basal delivery. 
        Using such abduction, we run the overdose query on UC2 (discussed in Section~\ref{sec:uc2}) to demonstrate that a~regular ``sunny day" narrative can be used to detect the overdose issue (implemented in \href{https://github.com/ovasicek/pca-pump-ec-artifacts/blob/master/narratives_and_queries/overdose-uc2-abduction.pl}{overdose-uc2-abduction.pl}).
        The query returns 4~different worlds of possible overdose.
        One of them, as an example, uses abduced values \code{initiallyP(vtbi_hard_limit_over_time(91/10 #=< V #< 101/10, 91))} with 3~query bindings, one of which is
        \code{141 #< T #=< 151} and \code{0 #< X #=< 1}.
        The max dose was abduced so that less than one bolus was allowed. 
        However, the bolus requested in UC2 was delivered, which caused an overdose during subsequent basal delivery.

        \subsubsection{An Overdose Error in the PCA Pump Requirements}
        \label{sec:overdose}
        The overdose issue is caused by enough boluses being delivered early in the timeline, particularly, close to the start of the pump.
        The cause of the issue is that there is no overdose protection measure specified for basal delivery.
        This is a~missing requirement that causes a~violation of a~critical safety property potentially causing harm to the patient depending on the overdose volume and the particular drug used.
        According to the authors of the PCA pump this is an unintentional omission.
        
        According to the requirement R5.2.0(5) (discussed as R5 in Section~\ref{sec:model-patient-bolus}), 
        when a~patient-requests a~bolus, the pump should reason about how much drug would be delivered within the max dose time window at the end of the currently requested bolus \emph{if} it was delivered.
        However, such reasoning only considers the contents of the max dose window in the past and does not consider what will follow in the future under normal operation.
        When a~patient requests a~bolus at a~time close enough to the start of the pump, then the max dose window starts at a~time smaller than the start time of the pump and, thus, includes a~period of zero drug delivery.
        With a~large enough max dose time window, enough boluses could be delivered to get close to the maximum safe dose.
        However, as the max dose window moves forward with time, the period of zero drug delivery is pushed out by the now in-progress basal delivery.
        And since basal delivery has no overdose protection measures, then it will keep running even if the maximum safe dose is exceeded.
        \\

        After fixing this issue by implementing the missing requirement (discussed in Appendix~\ref{apx:overdose-fix}, directly included in this arXiv version), the abductive query on UC2 no longer succeeds, meaning that an overdose was not found in the fixed model.
        The two versions of the model can be found in \href{https://github.com/ovasicek/pca-pump-ec-artifacts/blob/master/model-original.pl}{model-original.pl} and \href{https://github.com/ovasicek/pca-pump-ec-artifacts/blob/master/model-fixed.pl}{model-fixed.pl}.
        Of course, this is not a~sound proof of no overdose being possible---a~different overdose might be discoverable via different abducibles or narrative.

    \subsubsection{Results of Experiments with Validation of General Properties}
    \label{sec:overdose-comparison}
    
    Table~\ref{t:overdose-queries} shows results and execution times of querying overdose on variants of \EXC{}13 (discussed in Section~\ref{sec:ec13}) and of using abduction on UC2 (discussed in Section~\ref{sec:abductive_reasoning}).   
    Execution of the overdose queries takes much longer than the simulation queries from Table~\ref{t:simulation-queries} (minutes instead of seconds) due to the higher complexity of the overdose query.
    However, the abductive queries are the slowest ones due to the higher complexity of abduction in general but also due to the limitations of its current implementation in \scasp{}, discussed in Section~\ref{sec:incremental-abduction}.

\newcommand{\modellink}[1]{\href{https://github.com/ovasicek/pca-pump-ec-artifacts/blob/master/model-#1.pl}{#1}}
    \begin{table}
    \centering
    \small
    \caption{Overdose querying on \EXC{}13 and UC2}
    \vspace{2mm}
    \begin{tabular}{llllcr}
      \toprule 
                                                           & \textbf{Use Case Name}   & \textbf{Variant}             & \textbf{Model}      & \textbf{Result}& \textbf{Time (m)} \\
      \midrule
      \uclink{overdose-ec13b}{\EXC{}13b}                   & Maximum Safe Dose & patient bolus                & original            & OK             & 14.11             \\
      \uclink{overdose-ec13c}{\EXC{}13c}                   & Maximum Safe Dose & clinician bolus              & original            & OK             & 20.85             \\[.3em]        
      \multirow{2}{*}{\uclink{overdose-ec13a}{\EXC{}13a}}  & \multirow{2}{*}{Maximum Safe Dose}& \multirow{2}{*}{during basal}& original            & overdose       & 15.29             \\
                                                           & &                              & fixed               & OK             &  3.34             \\[.3em]
      \multirow{2}{*}{\uclink{overdose-uc2-abduction}{UC2}}& \multirow{2}{*}{Patient-Requested Bolus} & \multirow{2}{*}{abduction}   & original            & overdose       & 38.01             \\
                                                           & &                              & fixed               & OK             & 48.11             \\     
      \bottomrule
    \end{tabular}
    \label{t:overdose-queries}
    \end{table}

\section{Techniques Used to Empower s(CASP) Reasoning}

This section describes the techniques that we apply to avoid non-termination of
\scasp{} reasoning (Sections~\ref{sec:imp-ec-axioms}
and~\ref{sec:self-ending-trajectories}), the approach that we have proposed to
overcome limitations of \scasp{} abduction
(Section~\ref{sec:incremental-abduction}), and techniques proposed to
significantly improve reasoning performance (Sections~\ref{sec:cache}
and~\ref{sec:multi_run_approach}).


    \subsection{Improved Implementation of the Event Calculus Axioms}
    \label{sec:imp-ec-axioms}
    Our implementation of the BEC axioms (in
    \href{https://github.com/ovasicek/pca-pump-ec-artifacts/blob/master/ec_theory/bec_scasp-pca_pump.pl}{bec\_scasp-pca\_pump.pl})
    differs from the one by~\cite{arias-ec2022} in two aspects in
    order to avoid non-termination.
    First, inspired by \cite{gupta-train}, we use a~custom
    implementation of the \code{not} keyword.
    Namely, we implement negated predicates, such as \code{not_stoppedIn/3}, as
    simplified versions of the \emph{dual rules} that \scasp{} generates to
    compute the negated predicates.
    These simplified versions contain only the dual rules that are relevant for
    the intended evaluation of the negated predicates.
    Second, we introduce new predicates \code{can_initiates/2},
    \code{can_terminates/2}, \code{can_releases/2}, and \code{can_trajectory/4}.
    These are created by pre-processing the source code and
    introducing a~new fact \code{can_initiates(E,F).} for each fact and/or rule
    \code{initiates(E,F,T) :- some_body(E,F,T).} (and likewise for
    others).
    Our implementation of the BEC6 axiom (cf.
    Section~\ref{sec:scasp-asp-ec}) using these new predicates follows:
    %
    \begin{codeBox}
    \begin{lstlisting}[style=MySCASP]
holdsAt(Fluent, T2) :- T1 .<. T2, can_initiates(Event, Fluent),
  happens(Event, T1), initiates(Event, Fluent, T1), not_stoppedIn(T1, Fluent, T2).
    \end{lstlisting}
    \end{codeBox} 
    \noindent This construction is motivated by an observation that, in our
    experiments, proving the original predicate \code{initiates} first often
    leads to non-termination due to its sub-goals, while proving \code{happens}
    first often leads to non-termination and enlarges the search space due to
    the unconstrained \code{Event}.
    On the other hand, proving the sub-goal-free \code{can_initiates} first has
    proven reliable in avoiding non-termination and pruning the search space by
    constraining \code{Event}.
    A~similar approach was used by~\cite{shanahan_abductive}.

    \vspace{-0.2em} 
    \subsection{Modeling Non-termination-prone Self-ending Trajectories}\vspace{-0.1em} 
    \label{sec:self-ending-trajectories}
    The main challenge during the modeling of the PCA pump was non-termination caused by trajectories which we refer to as self-ending.
    A~trajectory, defined by a rule with a head \code{trajectory(F1,T1,F2,T2)}, starts when its control fluent~\code{F1} is initiated at some time \code{T1}, and the body of the rule then determines how the value of its continuous fluent~\code{F2} may be computed for any time \code{T2}, where \code{T1 .<. T2}, until~\code{F1} is terminated.
    The trajectory is \emph{self-ending} if~\code{F1} may be terminated at some time \code{T2} while the trajectory is active by some event~\code{E} that gets triggered when the value of \code{F2} satisfies a certain predefined condition.
    We call such an event a \emph{self-end event} of the given trajectory.
    In the PCA pump, almost all trajectories are self-ending, e.g., bolus deliveries terminate themselves based on how much drug they deliver.
    For example, the \code{trajectory(clinician_bolus_delivery_enabled(Duration),T1, total_drug_delivered(X),T2)}, defined in a~similar way as was shown in  Section~\ref{sec:model-patient-bolus}, is self-ending, and one of its self-end events is \code{clinician_bolus_halted_max_dose} because its trigger rule depends on the value of \code{total_drug_delivered} (the code below is simplified for space reasons, details can be found in \href{https://github.com/ovasicek/pca-pump-ec-artifacts/blob/master/model_sources/04-clinician_bolus_trajectory.pl\#L128}{04-clinician\_bolus\_trajectory.pl}):
    \vspace{-0.3em} 
    \begin{codeBox}
    \begin{lstlisting}[style=MySCASP]
happens(clinician_bolus_halted_max_dose, T2) :-
  initiallyP(vtbi_hard_limit_over_time(VtbiLimit, TimePeriod)),
  T1 .=. T2 - TimePeriod,   holdsAt(total_drug_delivered(TotalT1), T1),
  VtbiLimit .=. TotalT2 - TotalT1,   holdsAt(total_drug_delivered(TotalT2), T2).
    \end{lstlisting}
    \end{codeBox}
    \vspace{-0.3em} 
    \noindent The above rule will, currently, cause non-termination in \scasp{} (trace on \href{https://github.com/ovasicek/pca-pump-ec-artifacts/blob/master/narratives_and_queries/archived-test-run/loop-trace-for-clinician-halt-and-kvo-on-uc3a--outline.txt}{GitHub}).
    It is triggered when the amount of drug delivered within the max dose time window reaches the maximum allowed dose.
    The cause of the issue is that a~different trajectory, in this case representing KVO delivery, can be used to determine the value of \code{total_drug_delivered} while at the same time the start event of that trajectory, \code{KVO_started}, is triggered by the event \code{clinician_bolus_halted_max_dose}.
    This particular loop is created because KVO delivery is being considered as a~way to prove the value of \code{total_drug_delivered} at time~\code{T2}.
    However, clinician bolus delivery is the only type of delivery which can lead to success because only one delivery can be active at a~time, and if clinician bolus delivery was not active, then we would not need to reason about triggering its halt.
    \\
    
    To avoid this issue, we introduce a~new predicate \code{holdsAt/3} 
    and a~new axiom for EC.
    We use the new predicate to force the use of the right trajectory when proving the value of \code{total_drug_delivered} at \code{T2} at line~4 (defined above) by adding \code{clinician_bolus_delivery_enabled(_)} as a parameter to \code{holdsAt}. 
    The new axiom is the same as the BEC3 axiom, except for the addition of \code{Fluent1} as the third parameter:
    %
    \begin{codeBox}
    \begin{lstlisting}[style=MySCASP]
holdsAt(Fluent2, T2, Fluent1) :-
  can_trajectory(Fluent1, T1, Fluent2, T2),   can_initiates(Event, Fluent1),       
  happens(Event, T1),   initiates(Event, Fluent1, T1),
  trajectory(Fluent1, T1, Fluent2, T2),   not_stoppedIn(T1, Fluent1, T2).
    \end{lstlisting}
    \end{codeBox}
    %
    \noindent Specifying \code{Fluent1} ensures that only the trajectories controlled by that fluent will be considered when trying to prove \code{Fluent2}.
    In general, the \code{holdsAt/3} predicate should be used in self-end event trigger rules when one needs to prove the value of a~continuous fluent while its self-ending trajectory is active.
    
    \vspace{-0.3em} 
    \subsection{Abduction Using Incremental Refinement to Enforce Consistent Models}\vspace{-0.1em} 
    \label{sec:incremental-abduction}
    
    Abductive reasoning in \scasp{} can abduce a~different value of an abducible every time the abducible is reached in the reasoning tree.
    This is, however, unsuitable when some constant or a tuple of constants, representing, e.g., values of some parameters of the modeled system or of some scenario in which it is evaluated, is to be abduced.
    
    Since the above problem appears in our model, we have proposed its solution suitable for abducing numerical values.
    It is based on repeatedly refining the values abduced at different points in the reasoning tree---through repeatedly tightening constraints on possible values of the abducibles---until the same values are obtained everywhere (or the abduction fails).
    The solution consists of two phases; the first one follows:
    \vspace{-0.2em} 
    \begin{enumBox}
    \begin{enumerate}[itemsep=0pt,leftmargin=1.3em]
    \small
        \item Run an abductive query of $predicate(p_1, ..., p_n)$ where $p_i$ are variables to abduce.
        \item \textbf{For each} model $m$ produced by the query:
        \begin{enumerate}[itemsep=0pt]
            \item \textbf{For each} parameter $p_i$, the model $m$ will contain some number $y$ of value intervals $I_{1}^{p_i}, ...,I_{y}^{p_i}$ abduced at different points of the reasoning tree.
            \item \textbf{For each} $p_i$, compute the intersection $I^{p_i} = I_{1}^{p_i} \cap ... \cap I_{y}^{p_i}$.
            \item \textbf{If} $I^{p_i}$ is empty for some $p_i$ or if the exact combination of $I^{p_1}, ..., I^{p_n}$ has been seen before (globally across Step c) or if a~predefined cut-off depth has been reached, \textbf{then} discard $m$ and \textbf{end recursion}.
            \item \textbf{Else if} for all $p_i$, $I_{1}^{p_i} = ... = I_{y}^{p_i}$, \textbf{then} $m$ is a~\textbf{consistent model} with intervals of values $I^{p_1}, ..., I^{p_n}$ for $p_1, ..., p_n$. Add $(m, I^{p_1}, ..., I^{p_n})$ to the \textbf{result} and \textbf{end recursion}.
            \item \textbf{Else} make a~new query by restricting the abducible value of each parameter $p_i$ to $I^{p_i}$, and recursively perform Steps 1 and 2 using the new query.
        \end{enumerate} 
    \end{enumerate}
    \end{enumBox}
    \vspace{-0.2em} 

    The result of the first phase will be a set of models, each containing a tuple of intervals $I^{p_1}, ..., I^{p_n}$ of possible values of $p_1, ..., p_n$.
    However, in order to obtain one concrete witness of the result of the query, one cannot just take any tuple of values $v_1, ..., v_n$, where $v_i \in I^{p_i}$, since the values of the different parameters may depend on each other.
    Therefore, in the second phase of our solution, for each of the models, we proceed as follows.
    We select a value $v_n \in I^{p_n}$ and repeat the first phase with this value fixed, leading to new intervals $J^{p_1} \subseteq I^{p_1}, ..., J^{p_{n-1}} \subseteq I^{p_{n-1}}$.
    We then likewise gradually select and fix values for the parameters $p_{n-1}, ..., p_2$.
    For $p_1$, it is not needed to repeat the process since it is the last interval to pick a value from and, therefore, any choice will be valid.
    \\
    
    We use the above approach
    in Section~\ref{sec:abductive_reasoning} to abduce the initial value of a~constant fluent \code{initiallyP(vtbi_hard_limit_over_time(V,P))} with two variable parameters (the implementation is available in \href{https://github.com/ovasicek/pca-pump-ec-artifacts/blob/master/narratives_and_queries/incremental_abduction.sh}{incremental\_abduction.sh}).
    The described approach can find witnesses of a~property violation but, due to the cut-off bound in Step~2(c) of the first phase, it cannot guarantee that the property is not violated.
    It is also inefficient due to repeated executions which explore non-realistic parts of the state space, and, further, each execution is slower than our other experiments because it cannot use the experimental cache we introduced to optimize \scasp{} reasoning (discussed in Section~\ref{sec:cache}).
    However, despite the inefficiency, it was able to detect an error in the requirements specification (discussed in Section~\ref{sec:overdose}) in reasonable time.
    Introducing an efficient solution to this problem into \scasp{} is part of our future work.

    
    \subsection{Prototype Cache for Predicate Proof Results}
    \label{sec:cache}

    The runtimes presented in Tables~\ref{t:simulation-queries} and~\ref{t:overdose-queries} (except for abduction) were measured using \scasp{} version \version 
    under a~new, preliminary implementation of tabling that caches the first (un)successful evaluation of specific predicates.
    These predicates are selected using the \code{#table_once} directive, in a~similar way as mode-directed tabling, described by~\cite{tabling-modes-GuoG08-fixed} and~\cite{atclp-padl2019-fixed}, and implemented in several Prolog interpreters.
    Under this \emph{cache mode}, when one of the selected predicates fails to be proved as a~ground sub-goal
    of any rule, s(CASP) caches the failure (failure-tabling), and similarly, when the evaluation of the sub-goal succeeds, the success is cached.
    Subsequent attempts at proving the ground cached predicate will then use these results instead of attempting to prove it again.
    Note that since s(CASP) implements non-monotonic reasoning, the result is only valid while the current assumptions are valid---therefore, the result stays cached until current assumptions change. 
    \\

    Using the cache on our test suite, we have observed a~reduction of execution time by up to 95\,\% on individual test narratives with an overall average of 66\,\% across the whole test suite while still obtaining the same models (up to the cached parts of the proof tree). We believe that a~more sophisticated implementation of tabling, based on TCLP by~\cite{TCLP-tplp2019}, would increase the performance without losing soundness (note that for non-grounded sub-goals, we may lose other valid answers by storing only one answer, affecting the soundness of negated sub-goals) or completeness.
    
    In our experiments, we cached all EC predicates by including the file \href{https://github.com/ovasicek/pca-pump-ec-artifacts/blob/master/model_utils/cache.pl}{cache.pl} with the corresponding directives.
    Due to the nature of EC, proving anything at a~timepoint requires reconstructing the whole history from time zero to the given timepoint.
    Therefore, the history which had to be proven for the value of a~fluent at \code{T1} will potentially have to be re-proven again for \code{T2} where \code{T1 < T2}.
    The new cache is especially useful to prevent repeated \emph{failing attempts} to prove a~predicate such as \code{not_stoppedIn}.
    We found that even reasoning about simple narratives would attempt \emph{and fail} to prove \code{not_stoppedIn} many times for the exact same parameters.
    When using cache, the predicate will only fail to be proven once for each set of parameters.

    \subsection{Decoupling Triggers and Effects of Events into Multiple Executions}
    \label{sec:multi_run_approach}

    Based on our experiments, we believe that triggered events, especially the ones that can terminate a~trajectory, very significantly contribute to the solving complexity.
    This holds even for narratives in which such events are never actually triggered---because the reasoning keeps trying to prove their trigger due to their potential effects.
    To reduce the performance impact of triggered events, we propose below an approach that targets particularly those of such events that may only trigger once per narrative, such as certain alarms in the PCA pump.
    The idea is to use a \emph{multi-run reasoning} in which we decouple the trigger of such an event \code{E} from its effect.
    This is done by removing all effects of~\code{E} and moving them to a newly introduced event \code{E_EFFECT} instead (see an example in Appendix~\ref{apx:decopuling-example}, directly included in this arXiv version). 
    We then use one dedicated run to check whether \code{E} happens at some time \code{T} in the given narrative.
   If not, \code{E_EFFECT} will stay undefined in further reasoning.
   If~\code{E} does happen at~\code{T}, we introduce a fact \code{happens(E_EFFECT,T).}, which will then allow further reasoning to take the effect of \code{E} into account without having to reason about its trigger.
   \\
   
   We use the above approach for alarms related to the drug reservoir contents---\code{empty_reservoir_alarm} and \code{low_reservoir_warning} (defined in \href{https://github.com/ovasicek/pca-pump-ec-artifacts/blob/master/model_sources/08-drug_reservoir.pl\#L16}{08-drug\_reservoir.pl}).\footnote{Implemented in such a way that, for each narrative, we can choose to ignore the drug reservoir reasoning (when deemed not relevant), or to use the multi-run approach, or to re-enable the full (slow) reasoning.}
   Each of the alarms can only happen once in a narrative in response to the level of drug in the reservoir reaching a certain threshold since the reservoir cannot be refiled during a narrative.
   This approach was needed because implementing alarms related to the drug reservoir contents caused an unbearable slowdown for some narratives---the worst case in our test suite was a~slowdown from 6.7\,mins to 6.3\,hours, in a narrative where neither of the alarms happens.
   This is due to the fact that prior to implementing these alarms the PCA model only contained three triggered events that could terminate a trajectory, each terminating only one trajectory, and adding the new alarms introduced two new triggered events which together can terminate \emph{all} trajectories.
   Indeed, the effect of \code{low_reservoir_warning} is stopping any drug delivery and switching to KVO delivery and the effect of \code{low_reservoir_empty} is stopping the pump entirely. 
    
    We use the proposed approach for both the low reservoir warning and the empty reservoir alarm \emph{at the same time}
    for a~total of three executions (cf. \href{https://github.com/ovasicek/pca-pump-ec-artifacts/blob/master/narratives_and_queries/three_runs.sh}{three\_runs.sh}): the first query introduces a~new fact for the low reservoir, the second one introduces a~new fact for the empty reservoir while using the fact from the first query, and the third and final query considers both of the new facts.
    For a~narrative based on UC2 which reasons about both the low reservoir warning and the empty reservoir alarm, the three run approach takes 12 seconds while a~single run with full reasoning takes 35 minutes (cf. \href{https://github.com/ovasicek/pca-pump-ec-artifacts/blob/master/narratives_and_queries/empty_reserv-uc2-multirun-1.pl}{empty\_reserv-uc2-multirun-*} and \href{https://github.com/ovasicek/pca-pump-ec-artifacts/blob/master/narratives_and_queries/empty_reserv-uc2-onerun.pl}{empty\_reserv-uc2-onerun.pl}).
    We are experimenting with a~similar approach for incremental reasoning about all triggered events~as~future~work.

\section{Conclusions and Future Lines of Work}
\label{sec:conclusions}

Our work demonstrated that Event Calculus (EC) can be used to model the requirements specification of a~non-trivial, real-life cyber-physical system in \scasp{} 
and the reasoning involved can lead to discovering issues in the requirements while producing valuable evidence towards their validation.
Indeed, we have discovered a~violation of a~critical safety property in a~well-studied specification, acknowledged by its authors.

Our future work involves two directions.
The first includes improvements to \scasp{} by integration and efficient implementation of our abductive reasoning semantics,
improvements to prototype caching, 
and avoiding non-termination.
A common non-termination case is the ``toggle" scenario where a system toggles between two fluents affected by respective toggle events. 
A meta-reasoner in \scasp{} specialized to EC
would be more efficient and better at avoiding non-termination. 
The second direction involves software engineering to make our approach more general and practically usable, including
the replacement of unconstrained natural language requirements with structured languages like MIDAS, by \cite{midas}, for capturing requirements of industrial projects. 
This should provide enough structure and context to the requirements in order to enable a more general and at least semi-automated transformation of the requirements into EC, which would make our approach easier to adopt and use.

\bibliographystyle{tlplike}
\bibliography{scasp-pcapump}

\begin{thebibliography}{}

\bibitem[Arias and Carro, 2019a]{TCLP-tplp2019}
{\sc Arias, J.} {\sc and} {\sc Carro, M.} 2019a.
\newblock {D}escription, {I}mplementation, and {E}valuation of a {G}eneric {D}esign for {T}abled {CLP}.
\newblock {\em Theory and Practice of Logic Programming}, {\it 19}a, 3, 412--448.

\bibitem[Arias and Carro, 2019b]{atclp-padl2019-fixed}
{\sc Arias, J.} {\sc and} {\sc Carro, M.}
\newblock {Incremental Evaluation of Lattice-Based Aggregates in Logic Programming Using Modular {TCLP}}.
\newblock In {\em {Proc. of PADL'19~--~21st International Symposium on Practical Aspects of Declarative Languages}} 2019b, volume 11372 of {\em LNCS}, pp. 98--114. Springer.

\bibitem[Arias et~al., 2022]{arias-ec2022}
{\sc Arias, J.}, {\sc Carro, M.}, {\sc Chen, Z.}, {\sc and} {\sc Gupta, G.} 2022.
\newblock {Modeling and Reasoning in Event Calculus using Goal-Directed Constraint Answer Set Programming}.
\newblock {\em Theory and Practice of Logic Programming}, {\it 22}, 1, 51--80.

\bibitem[Arias et~al., 2018]{scasp-iclp2018}
{\sc Arias, J.}, {\sc Carro, M.}, {\sc Salazar, E.}, {\sc Marple, K.}, {\sc and} {\sc Gupta, G.} 2018.
\newblock {Constraint Answer Set Programming without Grounding}.
\newblock {\em Theory and Practice of Logic Programming}, {\it 18}, 3-4, 337--354.

\bibitem[Arnaud et~al., 2021]{cea-glossaries-and-process-algebras}
{\sc Arnaud, M.}, {\sc Bannour, B.}, {\sc Lapitre, A.}, {\sc and} {\sc Giraud, G.}
\newblock {Investigating Process Algebra Models to Represent Structured Requirements for Time-sensitive CPS}.
\newblock In {\em {Proc. of SEKE'21~--~The 33rd International Conference Software Engineering \& Knowledge Engineering}} 2021, Pittsburgh (virtual conference), United States.

\bibitem[Bhatt et~al., 2010]{hilite}
{\sc Bhatt, D.}, {\sc Madl, G.}, {\sc Oglesby, D.}, {\sc and} {\sc Schloegel, K.}
\newblock {Towards Scalable Verification of Commercial Avionics Software}.
\newblock In {\em Proc. of AIAA Infotech@Aerospace} 2010.

\bibitem[Crapo et~al., 2017]{ge-assert}
{\sc Crapo, A.}, {\sc Moitra, A.}, {\sc McMillan, C.}, {\sc and} {\sc Russell, D.}
\newblock {Requirements Capture and Analysis in ASSERT(TM)}.
\newblock In {\em Proc. of RE'17~--~25th International Requirements Engineering Conference} 2017. IEEE.

\bibitem[Gebser et~al., 2019]{clingo-paper}
{\sc Gebser, M.}, {\sc Kaminski, R.}, {\sc Kaufmann, B.}, {\sc and} {\sc Schaub, T.} 2019.
\newblock {Multi-shot ASP solving with clingo}.
\newblock {\em Theory and Practice of Logic Programming}, {\it 19}, 1, 27--82.

\bibitem[Gelfond and Lifschitz, 1988]{gelfond88:stable_models-fixed}
{\sc Gelfond, M.} {\sc and} {\sc Lifschitz, V.}
\newblock The {S}table {M}odel {S}emantics for {L}ogic {P}rogramming.
\newblock In {\em Proc. of 5th International Conference on Logic Programming} 1988, pp. 1070--1080.

\bibitem[Guo and Gupta, 2008]{tabling-modes-GuoG08-fixed}
{\sc Guo, H.-F.} {\sc and} {\sc Gupta, G.} 2008.
\newblock {S}implifying {D}ynamic {P}rogramming via {M}ode-directed {T}abling.
\newblock {\em Software: Practice and Experience}, {\it 38}, 1, 75--94.

\bibitem[Hall et~al., 2020]{midas}
{\sc Hall, B.}, {\sc Fiedor, J.}, {\sc and} {\sc Jeppu, Y.} 2020.
\newblock {Model Integrated Decomposition and Assisted Specification (MIDAS)}.
\newblock {\em INCOSE International Symposium}, {\it 30}, 1, 821--841.

\bibitem[Hatcliff et~al., 2019]{pcapump-paper}
{\sc Hatcliff, J.}, {\sc Larson, B.}, {\sc Carpenter, T.}, {\sc Jones, P.}, {\sc Zhang, Y.}, {\sc and} {\sc Jorgens, J.} 2019.
\newblock {The Open PCA Pump Project: An Exemplar Open Source Medical Device as a Community Resource}.
\newblock {\em SIGBED Rev.}, {\it 16}, 2, 8–13.

\bibitem[Larsen et~al., 2018]{timed-automata-uppaal}
{\sc Larsen, K.~G.}, {\sc Lorber, F.}, {\sc and} {\sc Nielsen, B.}
\newblock {20 Years of UPPAAL Enabled Industrial Model-Based Validation and Beyond}.
\newblock In {\em ISoLA'18~--~8th International Symposium on Leveraging Applications of Formal Methods, Verification and Validation} 2018, volume 11247 of {\em LNCS}. Springer.

\bibitem[Lifschitz, 2019]{asp}
{\sc Lifschitz, V.} 2019.
\newblock {\em Answer {Set} {Programming}}.
\newblock Springer.

\bibitem[Mueller, 2014]{mueller_book-fixed}
{\sc Mueller, E.~T.} 2014.
\newblock {\em {Commonsense Reasoning: An Event Calculus Based Approach}}.
\newblock Morgan Kaufmann.

\bibitem[Shanahan, 2000]{shanahan_abductive}
{\sc Shanahan, M.} 2000.
\newblock {An Abductive Event Calculus Planner}.
\newblock {\em The Journal of Logic Programming}, {\it 44}, 1-3, 207--240.

\bibitem[Varanasi et~al., 2022]{gupta-train}
{\sc Varanasi, S.~C.}, {\sc Arias, J.}, {\sc Salazar, E.}, {\sc Li, F.}, {\sc Basu, K.}, {\sc and} {\sc Gupta, G.}
\newblock {Modeling and Verification of Real-Time Systems with the Event Calculus and s(CASP)}.
\newblock In {\em In Proc. of PADL'22~--~Practical Aspects of Declarative Languages} 2022, volume 13165 of {\em LNCS}, pp. 181--190. Springer.

\end{thebibliography}

\newpage
\appendix

\section{Additional Details for Sections from the Paper}
\label{appendix:a}
This appendix contains further details which did not fit into the paper. In particular, a description of how we fixed an overdose issue in the PCA pump specification in Section~\ref{apx:overdose-fix}, an example of decoupling the trigger and effects for an event in Section~\ref{apx:decopuling-example}, and a~full version of a~table of experiments (instead of a~shortened version) in Section~\ref{apx:full-table}.

    \subsection{Fixing the Overdose Error in the PCA Pump Requirements}
    \label{apx:overdose-fix}
    Section~\ref{sec:overdose}
    discussed an overdose error that we have discovered in the requirements specification of the PCA pump.
    In this section, we discuss how we fixed the model in order to remove this issue.
    
    One option to fix the overdose issue is to introduce the missing overdose protection measure for basal delivery, as was the original intention defined in the exception cases
    in the specification.
    This solution would prevent harm to the patient, however, it could allow an early spike of boluses only to cause a~max dose warning during subsequent basal delivery, which leads to loss of therapy and requires the attention of a~clinician.
    Further, adding a~triggered event which can halt basal delivery based on the maximum safe dose, in a~similar way as what is already included for a~clinician-requested bolus, would cause a~significant increase in solving complexity and would turn basal delivery into a~self-ending trajectory.
    
    Therefore, we propose a~different, preemptive approach.
    When the pump reasons about denying a~bolus, it should consider that (at least) basal delivery will follow under normal operation by \emph{assuming a~max dose window full of basal delivery}.
    The actual amount of extra drug delivered via boluses is then added on top of the assumption.
    The overdose protection measure defined for clinician-requested boluses needs to be modified in a~similar fashion.
    In this way, there is no need for an overdose protection measure during basal delivery because the preemptive nature of handling boluses will not allow such situation, assuming that the max dose is defined such that basal delivery by itself (without any boluses) will not cause an overdose.
    Further, the preemptive denial of a~bolus should raise the max dose warning only for cases where the overdose would occur immediately after or during the bolus, not for cases where the overdose was preemptively prevented (i.e., only raise a~warning in cases when even the original model would).
    This approach is what our ``fixed'' model of the PCA pump currently uses.
    The two versions of the model can be found in \href{https://github.com/ovasicek/pca-pump-ec-artifacts/blob/master/model-original.pl}{model-original.pl} and \href{https://github.com/ovasicek/pca-pump-ec-artifacts/blob/master/model-fixed.pl}{model-fixed.pl}.

    \subsection{Example for the Decoupling of Event Triggers and Effects}
    \label{apx:decopuling-example}
    In 
    Section~\ref{sec:multi_run_approach}, we discussed a multi-run approach for mitigating increases in reasoning complexity caused by certain triggered events.
    The approach is based on decoupling the trigger of the event from the effects of the event.
    In this section, we provide a code demonstration of how the decoupling is done.
    Given some event \code{alarm}, which is a triggered event which can only be triggered up to once per narrative, it can have effects on some fluents via \code{initiates}/\code{terminates} and on other events via triggering their occurrence.
    The original code for the \code{alarm} event and its effects might look like this:
    \vspace{1em}
    \begin{codeBox}
    \begin{lstlisting}[style=MySCASP]
event(alarm).
happens(alarm, T) :- some_trigger_rule(T).  % event trigger
terminates(alarm, some_fluent, T).          % effect on fluents
happens(alarm_response_event, T) :-         % effect on other events
  happens(alarm, T).  
    \end{lstlisting}
    \end{codeBox}
    %
    \noindent In order to decouple the effects of the \code{alarm} event from its trigger, we replace the above code with the following code:
    \begin{codeBox}
    \begin{lstlisting}[style=MySCASP]
event(alarm).                               % original event
happens(alarm, T) :- some_trigger_rule(T).  % original event trigger

event(alarm_EFFECT).                        % new effect event
terminates(alarm_EFFECT, some_fluent, T).   % decoupled effect on fluents
happens(alarm_response_event, T) :-         % decoupled effect on other events
  happens(alarm_EFFECT, T).

happens(alarm_EFFECT,T):-                   % option to undo decoupling
  full_alarm_reasoning_enabled,
  happens(alarm,T).
    \end{lstlisting}
    \end{codeBox}
    %
    \noindent We introduce a new \code{alarm_EFFECT} event and move all the effects of the original \code{alarm} event to this new event (lines~4-7).
    The new event has no trigger rule, and the original event now has no effects.
    We also include an option to undo the decoupling in order to re-enable full reasoning (lines~9-11) via a configuration fact \code{full_alarm_reasoning_enabled}, which is not defined by default.
    
    For each narrative, we can either choose to ignore reasoning about the \code{alarm} event entirely (when it is deemed not relevant), or we can use the multi-run approach, or we can re-enable the full (slow) reasoning.
    For this particular example, the multi-run approach would consist of two runs.
    The first run would use the query \code{?- happens(alarm,T).}
    If the query fails, then the \code{alarm} event does not occur in the particular narrative and, thus, is not relevant.
    If the query succeeds, then we introduce a new fact \code{happens(alarm_EFFECT,t)}., where \code{t} is the binding for \code{T} from the result of the query.
    The second run of reasoning, then uses any query of interest while taking into account the new fact which might have been introduced in the previous run.

    In the case of decoupling multiple events, e.g., \code{E1} and \code{E2}, one needs to consider if one of the decoupled events has any impact on the occurrence of the other decoupled event.
    If the occurrence of \code{E1} influences the occurrence of \code{E2} in any way, then we need to use three runs in the multi-run approach.
    The first run queries the occurrence of \code{E1} and introduces the appropriate fact, the second run then queries the occurrence of \code{E2} while already considering the effects of \code{E1} via the new fact, and the third run then runs any query of interest while taking into account facts introduced by both of the prior runs.

    \subsection{Full Version of Table~\ref{t:simulation-queries}}
    \label{apx:full-table}
    Section~\ref{sec:simulation-comparison} presented Table~\ref{t:simulation-queries}, which was a~shortened version due to space constraints and due to limited value of presenting execution times of every UC/\EXC{}.
    We include the full version in Table~\ref{apx:t:simulation-queries-full}.
    %
    \begin{table}
        \centering
        \small
        \caption{Full results of simulation of relevant use cases and exception cases}
        \vspace{0.2em}
        \begin{tabular}{lllcr} \toprule
                                      & \textbf{Use Case Name}    & \textbf{Variant}         & \textbf{Result} & \textbf{Time(s.)} \\
          \midrule
          \uclink{uc1}  {UC1}         & Normal Operation          & no variants              & OK              &    1.69           \\
          \uclink{uc2}  {UC2}         & Patient-Requested Bolus   & no variants              & OK              &    3.17           \\
          \uclink{uc3a} {UC3a}        & Clinician-Requested Bolus & not suspend              & OK              &    2.84           \\
          \uclink{uc3b} {UC3b}        & Clinician-Requested Bolus & suspended and resumed    & OK              &   37.13           \\
          \uclink{uc7a} {UC7a}        & Resume Infusion           & after basal delivery     & OK              &    1.92           \\
          \uclink{uc7b} {UC7b}        & Resume Infusion           & after patient bolus      & OK              &    3.83           \\
          \uclink{uc7c} {UC7c}        & Resume Infusion           & after clinician bolus    & OK              &    3.56           \\
          \midrule
          \uclink{ec1}  {\EXC{}1}     & Bolus Request Too Soon    & no variants              & OK              &    2.72           \\
          \uclink{ec7a} {\EXC{}7a}    & Over-Flow Rate Alarm      & defined in \EXC{} step 1 & FAIL            &    1.87           \\
          \uclink{ec7b} {\EXC{}7b}    & Over-Flow Rate Alarm      & defined in \EXC{} step 1 & FAIL            &    1.87           \\
          \uclink{ec7c} {\EXC{}7c}    & Over-Flow Rate Alarm      & defined in \EXC{} step 1 & FAIL            &    2.17           \\
          \uclink{ec7d} {\EXC{}7d}    & Over-Flow Rate Alarm      & defined in \EXC{} step 1 & FAIL            &    4.62           \\
          \uclink{ec7e} {\EXC{}7e}    & Over-Flow Rate Alarm      & defined in \EXC{} step 1 & FAIL            &    1.53           \\
          \uclink{ec7f} {\EXC{}7f}    & Over-Flow Rate Alarm      & defined in \EXC{} step 1 & FAIL            &    4.24           \\
          \uclink{ec8a} {\EXC{}8a}    & Under-Flow Rate Alarm     & basal delivery           & OK              &    1.56           \\
          \uclink{ec8b} {\EXC{}8b}    & Under-Flow Rate Alarm     & defined in \EXC{} step 1 & FAIL            &    2.19           \\
          \uclink{ec8c} {\EXC{}8c}    & Under-Flow Rate Alarm     & defined in \EXC{} step 1 & FAIL            &    2.15           \\
          \uclink{ec9a} {\EXC{}9a}    & Pump Overheating          & during basal delivery    & OK              &    1.68           \\
          \uclink{ec9b} {\EXC{}9b}    & Pump Overheating          & during patient bolus     & OK              &    2.88           \\
          \uclink{ec9c} {\EXC{}9c}    & Pump Overheating          & during clinician bolus   & OK              &    2.31           \\
          \uclink{ec10a}{\EXC{}10a}   & Downstream Occlusion      & during basal delivery    & OK              &    1.67           \\
          \uclink{ec10b}{\EXC{}10b}   & Downstream Occlusion      & during patient bolus     & OK              &    3.15           \\
          \uclink{ec10c}{\EXC{}10c}   & Downstream Occlusion      & during clinician bolus   & OK              &    2.32           \\
          \uclink{ec11a}{\EXC{}11a}   & Upstream Occlusion        & during basal delivery    & OK              &    1.66           \\
          \uclink{ec11b}{\EXC{}11b}   & Upstream Occlusion        & during patient bolus     & OK              &    3.10           \\
          \uclink{ec11c}{\EXC{}11c}   & Upstream Occlusion        & during clinician bolus   & OK              &    2.35           \\
          \uclink{ec12a}{\EXC{}12a}   & Air-in-line Embolism      & during basal delivery    & OK              &    1.66           \\
          \uclink{ec12b}{\EXC{}12b}   & Air-in-line Embolism      & during patient bolus     & OK              &    2.91           \\
          \uclink{ec12c}{\EXC{}12c}   & Air-in-line Embolism      & during clinician bolus   & OK              &    2.31           \\
          \uclink{ec13a}{\EXC{}13a}   & Maximum Safe Dose         & during basal delivery    & FAIL            &   25.62           \\
          \uclink{ec13b}{\EXC{}13b}   & Maximum Safe Dose         & during patient bolus     & OK              &   31.61           \\
          \uclink{ec13c}{\EXC{}13c}   & Maximum Safe Dose         & during clinician bolus   & OK              &   53.45           \\
          \uclink{ec20a}{\EXC{}20a}   & Reservoir Low             & during basal delivery    & FAIL            &    1.99           \\
          \uclink{ec20b}{\EXC{}20b}   & Reservoir Low             & during patient bolus     & FAIL            &    7.95           \\
          \uclink{ec20c}{\EXC{}20c}   & Reservoir Low             & during clinician bolus   & FAIL            &    5.07           \\
          \uclink{ec21} {\EXC{}21}    & Reservoir Empty           & no variants              & OK              &   40.99           \\
          \bottomrule
        \end{tabular}
        \label{apx:t:simulation-queries-full}
    \end{table}

\end{document}